\documentclass[aps,prl,floats,floatsfix,twocolumn,superscriptaddress,showpacs]{revtex4-1}

\usepackage{dcolumn} % Align table columns on decimal point
\usepackage{bm} % bold math
\usepackage{amsthm}
\usepackage{latexsym}
\usepackage{float}
\usepackage{amsfonts}
\usepackage{amsmath}
\usepackage{amssymb}
\usepackage{color,graphicx}

\begin{document}

\title{Coherent Electronic Coupling in Atomically Thin MoSe$_{2}$}

\author{Akshay Singh}
\affiliation{Department of Physics, University of Texas, Austin, TX 78712, USA}
\author{Galan Moody}
\altaffiliation[Current address: ]{National Institute of Standards and Technology, Boulder, CO 80305, USA}
\affiliation{Department of Physics, University of Texas, Austin, TX 78712, USA}
\author{Sanfeng Wu}
\affiliation{Department of Physics, University of Washington, Seattle, WA 98195, USA}
\author{Yanwen Wu}
\affiliation{Department of Physics, University of Texas, Austin, TX 78712, USA}
\affiliation{Department of Physics and Astronomy, University of South Carolina, Colombia, SC 29208, USA}
\author{Nirmal J. Ghimire}
\affiliation{Department of Physics and Astronomy, University of Tennessee, Knoxville, Tennessee 37996, USA}
\affiliation{Materials Science and Technology Division, Oak Ridge National Laboratory, Oak Ridge, Tennessee, 37831, USA}
\author{Jiaqiang Yan}
\affiliation{Materials Science and Technology Division, Oak Ridge National Laboratory, Oak Ridge, Tennessee, 37831, USA}
\affiliation{Department of Materials Science and Engineering, University of Tennessee, Knoxville, Tennessee, 37996, USA}
\author{David G. Mandrus}
\affiliation{Department of Physics and Astronomy, University of Tennessee, Knoxville, Tennessee 37996, USA}
\affiliation{Materials Science and Technology Division, Oak Ridge National Laboratory, Oak Ridge, Tennessee, 37831, USA}
\affiliation{Department of Materials Science and Engineering, University of Tennessee, Knoxville, Tennessee, 37996, USA}
\author{Xiaodong Xu}
\affiliation{Department of Physics, University of Washington, Seattle, WA 98195, USA}
\affiliation{Department of Materials Science and Engineering, University of Washington, Seattle, WA 98195, USA}
\author{Xiaoqin Li}
\email{elaineli@physics.utexas.edu}
\affiliation{Department of Physics, University of Texas, Austin, TX 78712, USA}

\begin{abstract}
We report the first direct spectroscopic evidence for coherent electronic coupling between excitons and trions in atomically thin transition metal dichalcogenides, specifically monolayer MoSe$_2$.  Signatures of coupling appear as isolated cross-peaks in two-color pump-probe spectra, and the lineshape of the peaks reveals that the coupling originates from many-body interactions. Excellent agreement between the experiment and density matrix calculations suggests the formation of a correlated exciton-trion state due to their coupling.
\end{abstract}

\date{\today}
\maketitle

Atomically thin transition metal dichalcogenides (TMDs) have emerged as an interesting class of two-dimensional materials due to their unique optical properties.  For example, some materials (e.g. MoS$_2$) exhibit a crossover from an indirect-to-direct gap semiconductor as the material thickness is reduced to one atomic layer \cite{MakPRL2010}, consequently enhancing the photoluminescence efficiency \cite{Splendiani2010}.  Valley-specific optical selection rules \cite{MakNN2012,ZengNN2012,CaoNC2012} dictated by time reversal and spatial inversion symmetries in monolayer structures have been observed, and such selection rules have been electrically controlled in bilayers \cite{WuNP2013}.  These attributes make TMDs particularly intriguing for tunable and flexible valleytronic and photonic devices \cite{WangNN2012}.

A striking feature in the linear optical spectra of atomically thin TMDs is the presence of pronounced exciton resonances (Coulomb-bound electron-hole pairs), even at room temperature. The large binding energy of excitons, estimated to be hundreds of meV \cite{Rama2012,Lambrecht2012}, as well as tightly-bound trions (an exciton bound with an extra electron or hole) \cite{MakNM2013,RossNC2013} arises from reduced dielectric screening in the thickness direction of the film \cite{Jena2007}. It is reasonable to speculate that enhanced Coulomb interactions responsible for the large exciton and trion binding energies should also lead to strong coherent coupling among these quasiparticles. The presence or absence of coupling between excitonic states in conventional semiconductors is known to significantly influence energy transfer \cite{Snoke2002}, photon emission statistics \cite{Stevenson2006}, and quantum-logic operations \cite{LiScience2003} in quantum wells, wires, and dots. It is particularly imperative to understand the properties of these intrinsically many-body states in TMDs as they are inevitably present in all optical devices such as photovoltaics \cite{Britnell2013,Fontana2013}, detectors \cite{Sanchez2013}, and modulators \cite{Yin2012,Newaz2013}.  The existence and manifestation of electronic coupling is challenging to study experimentally since nonlinear spectroscopy techniques are required. Previous experiments thus far have provided limited information on the effects of phase-space filling, exciton-carrier broadening effects \cite{WangPRB2012,ShiACS2013}, inter-excitonic scattering \cite{SimPRB2013}, valley relaxation dynamics \cite{Mai2014}, and biexciton formation \cite{Sie2013} in TMDs.

In this Letter, we provide clear evidence for coherent exciton-trion interactions in monolayer MoSe$_2$ using two-color ultrafast pump-probe spectroscopy. In a high quality sample with spectrally well-resolved exciton and trion resonances, spectroscopic signatures for electronic coupling between these two types of quasiparticles are isolated as cross-peaks in a two-dimensional map when the pump and probe wavelengths are tuned independently through these transitions. While incoherent population relaxation may partially contribute to the coupling peaks, density matrix calculations reveal that the unique lineshape of the peaks arises from coherent exciton-trion many-body interactions.  Coherent coupling among optically excited quasiparticles in atomically thin materials is likely a common phenomenon due to lack of screening in the thickness direction.  Understanding such coupling is a prerequisite in controlling material properties at the electron level.  The results presented here add to the rich phenomena arising from many-body interactions in these intriguing quantum materials and will likely motivate continued efforts towards the development of a complete microscopic theory.

\begin{figure}[h]
\centering
\includegraphics[width= 0.85\columnwidth]{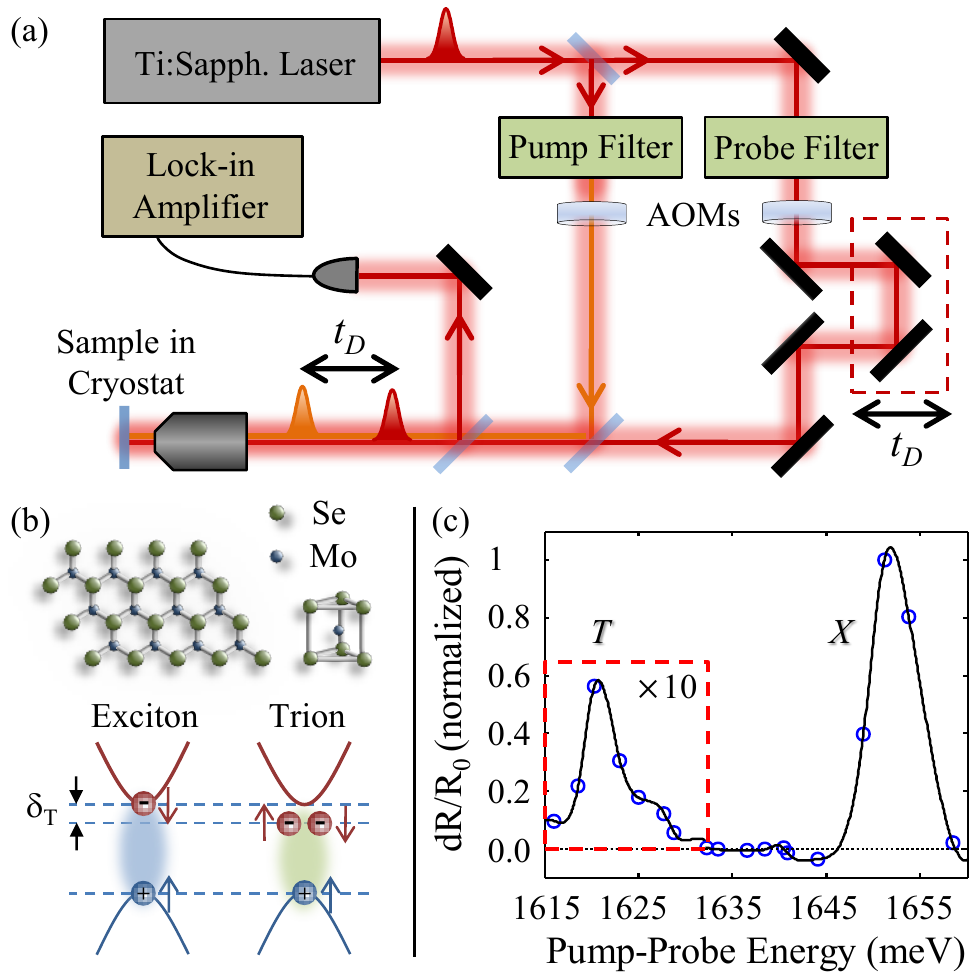}
\caption{(Color online) (a) Schematic diagram of the two-color pump-probe setup. (b) The hexagonal lattice structure of monolayer MoSe$_2$. Optical excitation generates either excitons ($X$) or trions ($T$) red-shifted from the exciton by an energy $\delta_{\textrm{T}}$. (c) Degenerate pump-probe spectrum (points) for delay $t_D = 0.7$ ps.  The curve serves as a guide-to-the-eye.}
\label{fig1}
\end{figure}

The experimental setup is illustrated in  Fig. \ref{fig1}(a).  The output of a mode-locked Ti:sapphire laser (90 MHz repetition rate and $15$ nm full-width at half-maximum bandwidth) is split into two beams.  The spectrum of each beam is modified independently using a grating-based pulse shaper, producing $\sim 1$ ps pulses (corresponding to $\sim 2$ nm bandwidth full-width at half-maximum).  The pump and probe beams are combined collinearly and are focused to a $\sim 3$ $\mu$m spot size on the sample. The two beams are cross-linearly polarized to suppress scattering from the pump into the detection optics. The average power of both beams is kept below $10$ $\mu$W to ensure that the signal remains in the $\chi^{\left(3\right)}$ regime.  The pump-induced change in the probe reflectivity is recorded using a lock-in amplifier while the pump and probe wavelengths and delay, $t_D$, are systematically varied.  Monolayer MoSe$_2$ was obtained through mechanical exfoliation onto 285-nm-thick SiO$_2$ on an $n$-doped silicon substrate.  Isolation of a single layer was verified through optical and atomic force microscopy, and the sample was mounted in a closed-loop cryostat at a temperature of 13 K for the optical spectroscopy experiments.

In MoSe$_2$, molybdenum and selenium atoms form a two-dimensional hexagonal lattice with a trigonal prismatic coordination (Fig. \ref{fig1}(b)).  At the K points in the first Brillouin zone, monolayer MoSe$_2$ has a direct bandgap with similar electron and hole band curvatures \cite{MakPRL2010,Li2007}.  Monolayer TMDs are unique in that broken inversion symmetry and strong spin-orbit effects couple the valley and spin degrees of freedom at the $\pm$K points.  Because cross-linear polarization is used to suppress pump scattering in the experiment, we do not distinguish between the valley or spin degrees of freedom of excitons or trions and instead focus on coherent coupling mechanisms between these quasiparticles.

\begin{figure}[h!]
\centering
\includegraphics[width=0.7\columnwidth]{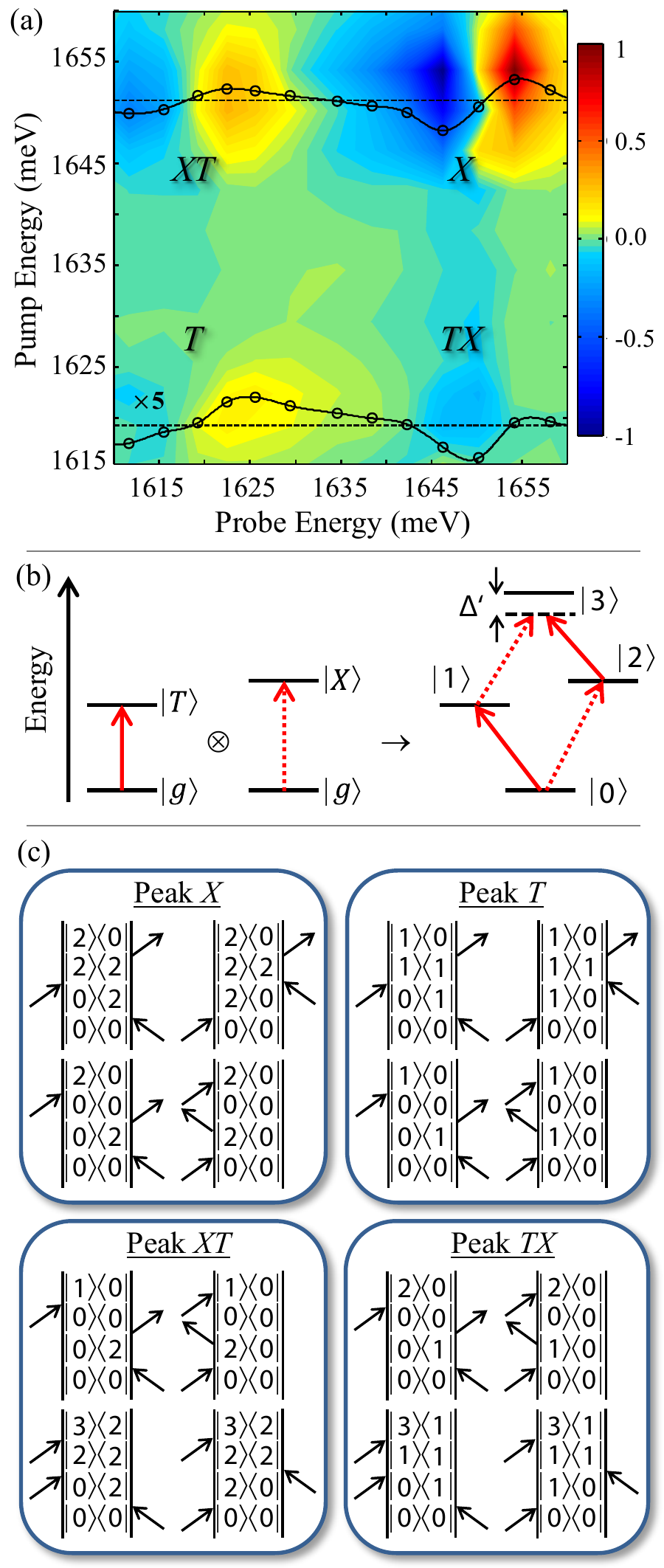}
\caption{(Color online) (a) Two-dimensional spectrum featuring exciton ($X$) and trion ($T$) peaks and their coupling ($XT$ and $TX$).  Differential probe spectra (points) are shown for the pump resonant with the exciton (top) and trion (bottom).  The curves serve as a guide-to-the-eye. (b) Two-independent two-level systems representing the exciton ($\left|g\right> \rightarrow \left|X\right>$) and trion (($\left|g\right> \rightarrow \left|T\right>$) transitions. An equivalent four-level scheme with ground ($|0\rangle$), trion ($|1\rangle$), exciton ($|2\rangle$) and exciton-trion ($|3\rangle$) states. (c) Double-sided Feynman diagrams representing the quantum  pathways relevant for the experiment.}
\label{fig2}
\end{figure}

We first perform a degenerate pump-probe experiment in which both beams are derived from the same pulse shaper.  The degenerate pump-probe spectrum, shown in Fig. \ref{fig1}(c) for $t_D = 0.7$ ps, was used to identify exciton ($X$) and trion ($T$) resonances at $1651$ meV and $1619$ meV, respectively. The transition energies and trion binding energy of $\delta_{\textrm{T}} \approx 30$ meV are consistent with values obtained from photoluminescence spectra of similarly-prepared samples \cite{RossNC2013}.  One-dimensional pump-probe spectra, however, cannot provide information regarding coherent coupling between resonances.  Two-color pump-probe spectroscopy overcomes this limitation, as demonstrated by the normalized spectrum shown in Fig. \ref{fig2}(a).  The spectrum was acquired for a delay $t_D = 0.7$ ps as a compromise between avoiding ambiguities in the pump-probe time ordering and minimizing contributions from incoherent population transfer.  The spectrum features four peaks -- two peaks corresponding to the exciton ($X$) and trion ($T$), a cross-peak when pumping at the exciton and probing at the trion ($XT$), and vice versa ($TX$).  The appearance of cross-peaks is an unambiguous sign of exciton-trion coupling.  A differential probe spectrum for resonant pumping at the exciton and trion is shown by the upper and lower horizontal slices, respectively, where the curves serve as a guide-to-the-eye.  Differences in the line shape of the $XT$ and $TX$ peaks suggests that the coupling may not be simply due to incoherent population transfer and instead originates from exciton-trion many-body interactions.

To elucidate the coupling mechanism, we use a phenomenological model to simulate a four-level system in the excitation picture.  Without making any assumption about whether exciton and trion resonances are coupled, one can map two two-level systems into a four-level system through a Hilbert space transformation, as illustrated in Fig. \ref{fig2}(b).  State $|0\rangle$ refers to the crystal ground state in the absence of any optical excitation; states $|1\rangle$ and $|2\rangle$ represent the trion and exciton, respectively; and state $|3\rangle$ represents simultaneous excitation of the exciton and trion.  The dynamics of the system, including population relaxation and dephasing, can be described by the density matrix, whose elements are related to the nonlinear signal via the optical Bloch equations (OBEs) \cite{Scully1997}.  We solve the OBEs perturbatively up to third order in the excitation field to calculate the nonlinear signal detected in the experiment.  The total nonlinear response can be generated using a sum-over-states expression derived from double-sided Feynman diagrams \cite{Mukamel1995}, illustrated in Fig. \ref{fig2}(c), each of which represents a quantum mechanical pathway that contributes to the signal.  For each diagram, the bottom two arrows indicate interaction of the pump field with the sample and the top arrow is interaction of the probe field after a delay $t_D$.  From each diagram, one can generate an expression for the perturbative evolution of the density matrix \cite{SM}.

The four-level energy scheme admits the inclusion of many-body interactions in a simple and intuitive manner.  Since the lower transitions ($|0\rangle \leftrightarrow |1\rangle$ and $|0\rangle \leftrightarrow |2\rangle$) are excited by the optical field to first order in perturbation theory, while the upper transitions ($|1\rangle \leftrightarrow |3\rangle$ and $|2\rangle \leftrightarrow |3\rangle$) contribute to third order only if the lower transitions have been excited, the electronic and optical properties of the upper transitions dictate the interaction strength \cite{Bott1993}.  Specifically, we consider excitation-induced shift (EIS) and excitation-induced dephasing (EID) effects, which have been used to explain coherent exciton coupling in semiconductor quantum wells \cite{Nardin2013,Moody2013_1} and quantum dots \cite{Kasprzak2011,Moody2013}.  EIS and EID correspond to the real and imaginary part of the renormalization energy when interaction effects are considered, thus both effects must appear simultaneously, in principle.  In the calculations, EIS is modeled by breaking the energy equivalence of the lower and upper transitions through a shift, $\Delta'$, of state $|3\rangle$.  EID is introduced by enhancing the dephasing rate of either or both of the upper transitions with respect to the equivalent lower transition by an amount $\gamma'$.

To illustrate how these effects are included in the calculation, we explicitly write down an expression for the perturbative evolution of one density matrix component.  Specifically, the quantum pathway corresponding to pumping at the trion (state $|1\rangle$) and probing at the exciton (state $|2\rangle$) via excitation of the exciton-trion state ($|3\rangle$), represented by the bottom two Feynman diagrams for peak $TX$ in Fig. \ref{fig2}(c), is described by the evolution of the third-order polarization $\rho^{\left(3\right)}_{13}$:
\begin{equation}
\begin{split}
\dot{\rho}^{\left(3\right)}_{13} = \left[-i\left(\omega_{20} - \frac{\Delta'}{\hbar}\right) - \left(\gamma_{20} + \gamma'\right)\right]\rho^{\left(3\right)}_{13} + \frac{i\mu_{13}}{2\hbar}\hat{E}\rho^{\left(2\right)}_{11},
\end{split}
\label{ExampleOBE}
\end{equation}
where the dipole moment, resonance energy, and dephasing rate of the transition are given by $\mu_{13}$, $\hbar\omega_{13} \equiv \hbar\omega_{20} - \Delta'$ and $\gamma_{13}\equiv\gamma_{20}+\gamma'$, respectively, $\hat{E}$ is the electric field, and $\Delta'$ and $\gamma'$ are the EIS and EID parameters for this pathway.  We describe how we have chosen the parameters for all pathways in detail in the supplemental material \cite{SM}. However, the precise values \textit{do not} affect the line shape qualitatively. Therefore, by carefully analyzing the lineshape of each peak in the 2D map, the role of coherent many-body interactions can be identified as explained in the following.  We emphasize that analyzing all peaks in the 2D spectrum simultaneously is essential, since the lineshape of each peak is sensitive to an arbitrary phase shift between the reflected signal and probe.  However, this phase shift is nearly constant across the entire 2D spectrum \cite{Phase}; thus the \textit{relative} amplitude and lineshape of the peaks provides critical information that enables us to distinguish between the coupling mechanisms.

\begin{figure*}[t]
\centering
\includegraphics[width=1.8\columnwidth]{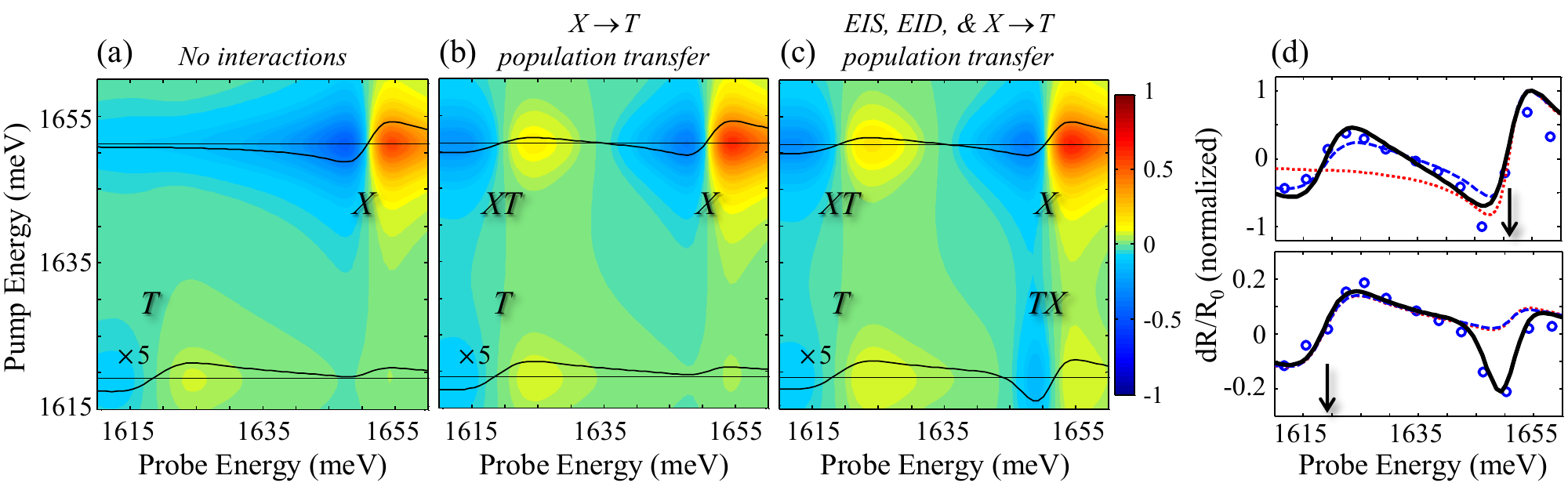}
\caption{(Color online) Simulated two-color pump-probe spectra (a) without interactions and with interactions via (b) incoherent exciton $\rightarrow$ trion population transfer and (c) EIS, EID, and exciton $\rightarrow$ trion transfer. (d) Comparison between experimental (points) and simulated (curves) differential probe spectra for resonant pumping at the exciton and trion resonances (indicated by vertical arrows).  The red (dotted), blue (dashed), and black (solid) curves correspond to the simulations in (a), (b), and (c), respectively.}
\label{fig3}
\end{figure*}

To elucidate interaction effects in the measured 2D spectrum, we present and discuss results of simulated spectra in three different scenarios. First consider the simulated spectrum with no exciton-trion interactions (Fig. \ref{fig3}(a)) for which no cross-peaks are present. The absence of interactions is modeled by using the same parameters (dipole moment, dephasing rate, and resonance energy) for the corresponding lower and upper transitions.  In this case, the quantum pathways responsible for coupling between the lower transitions are completely cancelled by pathways involving the upper transitions.  Thus the remaining non-zero quantum pathways are those associated with the individual exciton and trion resonances, and no cross-peaks appear in the spectrum \cite{NoInteractionPeak}.

In the case that only population relaxation between the exciton and trion states is considered, the 2D map (Fig. \ref{fig3}(b)) features highly asymmetric cross-peaks. Population transfer is modeled by including additional decay and source terms to the Feynman diagrams.  We note that the exciton and trion relative amplitudes are related not only to quantities that characterize these transitions (dipole moments and dephasing rates), but also to the background charge density determined by the unintentional $n$-doping in the material \cite{DopingType}.  Therefore, we examine the amplitude ratio between $TX/T$ and $XT/X$ as the key parameter for evaluating the role of incoherent population relaxation. For pumping resonantly with the exciton, formation of the $XT$ cross-peak requires the capture of an extra electron. This capture process is energetically favorable, thus, incoherent population relaxation may lead to an appreciable amplitude of the cross-peak (peak $XT$ in Fig. \ref{fig3}(b)) if the background charge density is not too low. Conversely, when pumping resonantly with the trion, only dissociation of the trion into an exciton and a free electron may lead to the cross-peak $TX$. This process requires additional energy from other mechanisms, such as annihilation of a phonon with an energy equal to the trion binding energy $\delta_T \approx 30$ meV. The phonon population density at this energy at a temperature of $13$ K is minimal. Therefore, one would expect a small amplitude ratio between $TX$ and $T$. Clearly, incoherent population alone cannot explain the large amplitude of the $TX$ peak in Fig. \ref{fig2}(a). Additionally, our simulation demonstrates that incoherent population transfer leads to the same lineshape for $TX$ and $XT$ (not shown), which is inconsistent with the experiment.

Finally, the role of coherent coupling mechanisms is examined. The term ``coherent" is used to distinguish these mechanisms from incoherent population relaxation processes. In the simulated spectrum shown in Fig. \ref{fig3}(c), in which both EIS and EID are included in addition to $X \rightarrow T$ population transfer, excellent agreement with the measured spectrum is obtained.  To make a more qualitative comparison between the simulation and experiment, two horizontal cuts are taken through the 2D maps and the results are shown in Fig. \ref{fig3}(d). The pump wavelength (indicated by the arrow) was tuned to the exciton and trion resonances, respectively, in the top and bottom panels.  The curves are results of the simulation using a fitting procedure with a limited number of parameters (see supplemental material).  Best agreement is obtained when including EIS and EID effects (solid curves), particularly for the $TX$ peak.  This fitting suggests a binding energy for the exciton-trion correlated state of $\Delta' = 4 \pm 1.5$ meV for the excitation conditions used in the experiment \cite{EISModel}.  We emphasize that the line shape of peak $TX$--a negative peak rather than the dispersive lineshape for the other three peaks--can only be reproduced by including the EIS effect. This distinct lineshape cannot be reproduced through the inclusion or adjustment of any other parameters.

These results demonstrate that reduced dielectric screening in TMDs results in enhanced long-range Coulomb interactions between quasiparticles compared to conventional semiconductors.  For example, $\Delta' \approx 4$ meV observed here is at least an order of magnitude larger compared to exciton-trion coupling in a 20-nm-wide $n$-doped CdTe/CdMgTe quantum well \cite{Moody2013_1}.  Strong exciton-trion coherent coupling makes TMDs an excellent platform for opto-electronic devices designed for a variety of coherent spin phenomena and conditional control schemes for quantum information applications.  We anticipate that complete microscopic many-body calculations will provide additional insight into the effects of Coulomb correlations on quasiparticle interactions in TMDs \cite{Chemla2001}.  Such efforts can be facilitated by advanced spectroscopic techniques such as coherent multi-dimensional spectroscopy, which can isolate specific quantum pathways \cite{Li2013,Stone2009}.

The work at UT was supported by the NSF (DMR-0747822), the Welch Foundation (F-1662), ARO-W911NF-08-1-0348, AFOSR-PECASE: FA9550-10-1-0022, and the Alfred P. Sloan Foundation.  The work at UW was supported by U.S. DoE, BES, Materials Sciences and Engineering Division (DE-SC0008145).  The work at ORNL was supported by U.S. DoE,  Office of Basic Energy Sciences, Materials Sciences and Engineering Division.

\bibliography{RefList}{}

\begin{thebibliography}{41}
\bibitem [{Pha()}]{Phase}%
  \BibitemOpen
  \href@noop {} {\bibinfo  {journal} {A phase shift due to the
  wavelength-dependent reflection coefficient is included, resulting in a
  change of 0.15 radians over the wavelength range of our measurement}\
  }\BibitemShut {NoStop}%
\bibitem [{NoI()}]{NoInteractionPeak}%
  \BibitemOpen
\bibfield  {journal} {  }\href@noop {} {\bibinfo  {journal} {The apparent small
  peak in Figs. 3(a) and 3(b) at the $TX$ peak position is due to spectral
  overlap between $X$ and $T$ resonances}\ }\BibitemShut {NoStop}%
\end{thebibliography}

\end{document}